\documentclass[12pt]{iopart}
\usepackage{bm,amssymb,graphicx}
\usepackage{iopams}
\usepackage[]{xcolor}
\usepackage{tikz}

\newcommand{\medio}[1]{\left\langle #1\right\rangle}
\newcommand{\eqref}[1]{(\ref{#1})}

\newcommand{\corr}[1]{#1}
\newcommand{\ehg}[1]{#1}

\begin{document}

\title[Nonlocal birth-death competitive dynamics with volume exclusion]{Nonlocal
birth-death competitive dynamics with volume exclusion}

\author{Nagi Khalil, Crist\'obal L\'opez, and Emilio Hern\'andez-Garc\'{\i}a}

\address{IFISC (CSIC-UIB), Instituto de F\'{\i}sica Interdisciplinar y Sistemas Complejos,
Campus Universitat de les Illes Balears,
E-07122, Palma de Mallorca, Spain.} \ead{nagi@ifisc.uib-csic.es}

\begin{abstract}
A stochastic birth-death competition model for particles with
excluded volume is proposed. The particles move, reproduce, and
die on a regular lattice. While the death rate is constant, the
birth rate is spatially nonlocal and implements inter-particle
competition by a dependence on the number of particles within a
finite distance. The finite volume of particles is accounted
for by fixing an upper value to the number of particles that
can occupy a lattice node, compromising births and movements.
We derive closed macroscopic equations for the density of
particles and spatial correlation at two adjacent sites. Under
different conditions, the description is further reduced to a
single equation for the particle density that contains three
terms: diffusion, a linear death, and a highly nonlinear and
nonlocal birth term. Steady-state homogeneous solutions, their
stability which reveals spatial pattern formation, and the
dynamics of time-dependent homogeneous solutions are discussed
and compared, in the one-dimensional case, with numerical
simulations of the particle system.
\end{abstract}



\vfill \today

\maketitle

\section{Introduction \label{sec:1}}

Birth-death models are a type of individual-based models (IBMs)
inspired by chemical, physical, and biological systems, where
the random motion of the particles is coupled to a reactive
dynamics \cite{Grimm}.
The usual reaction terms may include annihilation, reproduction
or coalescing processes which lead in general to a nonconserved
total number of particles. Many recent works have used IBMs to
describe biological systems, and have shed light on the
formation of clusters for both non-interacting
\cite{Young,Serva} and interacting \cite{helo, Els} birth-death
systems of ideal particles of vanishing size. However, the role
of the size or excluded volume, though widely studied in other
contexts
\cite{alwa57,hamc06,paza10,khcafipico14,doed88,madi99,krmasa14,codeensa10}
and its large importance for biological systems (dynamics of
microorganisms \cite{ab79,mamcle04,deaubagr09,plsi12,dymaba12},
animal flocks \cite{grchtu01,grch04,bilosp13,elwigo15},
chemotaxis \cite{thcopaybbo12,post14,limapaca15}), \corr{has
not received much attention. \ehg{See some exceptions in
\cite{maagbabema17} and references therein.} }

Many works show the influence of excluded volume effects on
randomly moving particles
\cite{VanBeijeren,Bruna,ruthbaker,burger2010,chikrma14}, both
considering the size of the particles itself or rather the
maximum number of particles allowed at each node of a lattice.
These analyses have been performed for the discrete particle
dynamics and for its continuum  description in terms
of a density or concentration field. It remains unknown what
happens when interactions modulating a birth-death dynamics
enter into play. The effects could be nontrivial, because
generally the spatial range of the interactions affecting the
birth-death processes would be different from the particle size
which gives the excluded-volume effect. This is the focus of this paper. We try to
answer the following questions: what is the effect of
considering the volume of the particles in a particular
interacting birth-death model? what is the macroscopic
description of this system? does volume-exclusion affect the
birth and death rates? what is its role on the clustering of
individuals?

We address them for the case in which the interactions between
the individuals are of competitive type. To this end we present
a stochastic model of a system of bugs living on a regular
lattice, adapting a previous birth-death model introduced
\cite{helo} to simulate the dynamics of competing individuals
through nonlocal spatial interactions. The new ingredient of
the model, concerning the finite volume of the particles, is
controlled by tuning the number of allowed particles per node,
as in the generalized exclusion process \cite{chikrma14}. The
stochastic model allows  to incorporate the main ingredients of
the processes, and to investigate the range of validity of an
eventual macroscopic description. In general, the latter
description involves a hierarchy of equations for the moments
of the multiparticle probability distribution (including the
density of particles and their correlations). \corr{The
infinite hierarchy is truncated by proposing a factorization of
the three--node correlations}. If the number of allowed
particles in a node is infinity, i.e. there is no volume
exclusion, and fluctuation correlations are small enough, we
recover the macroscopic equation derived in previous works with
different approaches \cite{helo}. In the case of full volume
exclusion, i.e. at most one particle allowed per site, we
analyze the conditions under which correlations can be
approximated to obtain a closed equation for the average
density field. This density equation is our main result and
presents two important features: a) there is no effective
diffusivity coming from the volume exclusion (this is already
the case for particle-conserving dynamics with single-particle
maximum occupation, as pointed in \cite{chikrma14}), and b) a
cubic density term in the reaction part accounts for this. We
analyze this equation, \ehg{and in the one-dimensional case we
obtain} the phase diagram where the different solutions are
shown and compared with the numerical simulation of the
stochastic particle dynamics.

The outline of the paper is the following. In the next section
we introduce the individual-based model. In Sec.
\ref{sec:macro} we derive the macroscopic description. In Sec.
\ref{sec:4} we analyze the homogeneous solutions of the density
equation and make a linear  stability analysis  to see when
spatial patterns arise. In Sec. \ref{sec:5} we
compare the theoretical results with the numerical simulations
of the stochastic particle model. Finally, in Sec. \ref{sec:6}
we discuss our results.

\section{Individual based model} \label{sec:ibm}
\subsection{Model}

In brief, we consider a model of competing finite-size
organisms that randomly move (diffuse) in space, and that may
die or reproduce. In this last case the newborn is located in a
neighboring position if available. Competition is introduced in
the system via the birth rate, so that the probability of
reproduction decreases with the number of individuals within a
given neighborhood of given radius $R$.

More in detail, the system is an ensemble of identical
particles living on a $d$--dimensional hypercubic lattice
$\Sigma$, embedded in $\mathbb{R}^d$, with periodic boundary
conditions. The total number of nodes is $N$ and the lateral
length of the system is $L$. Any possible configuration of the
system is given by the set $S\equiv \{s_i\}_{i=1}^N$ of
occupation numbers $s_i\in\{0,\dots,\sigma_i\}$, where
$\sigma_i$ is the maximum number of particles that node $i$ can
have. Given a node $i$ we define two sets of neighboring nodes,
characterizing the two types of interactions occurring in the
system: $N_i$ is the set of first neighbors of node $i$
including itself. In our hypercubic lattice each of the sets
$N_i$ contains $2d+1$ nodes. Excluded volume interactions will
be implemented in terms of this neighborhood $N_i$. The second
set is $P_i$ the set of nodes experiencing competition with
$i$, i.e. the sites whose distance to $i$ is smaller or equal
to $R$.

The dynamics of the system is given by a continuous--time
Markov chain involving three independent processes:
\begin{itemize}
\item Deaths: There is a constant probability rate for
    death $r_d$ for a particle at any node, hence the total
    death rate at node $i$ is
  \begin{equation}
    \label{eq:1}
    \pi_d(s_i)=r_ds_i.
  \end{equation}
\item Births: A particle at node $i$ gives birth \corr{to} another
    one at $j\in N_i$ with probability rate $(r_b-\alpha
    m_i)_\theta(\sigma_j-s_{j})$, where $r_b$ is a positive
    constant (the actual birth rate if the particle has
    only one empty accessible site and there is no
    competition), $\alpha>0$ is a constant that accounts
    for the effect of resource competition, $m_i=\sum_{j\in
    P_i}s_j$ is the total number of organisms in the set
    $P_i$, and the subscript $\theta$ of $(r_b-\alpha
    m_i)_\theta$ is a shortcut of $(r_b-\alpha
    m_i)\theta(r_b-\alpha m_i)$ with $\theta$ being the
    Heaviside function, insuring that this rate does not
    become negative. Note that the particle birth rate into
    site $j$ is proportional to, $\sigma_j-s_j$, i.e. the
    available capacity of this site to hold more particles,
    an implementation of the excluded-volume effect. The
    total probability rate of births at node $i$ is then
  \begin{equation}
    \label{eq:2}
    \pi_b(s_i,S)=(\sigma_i-s_i) \sum_{j\in N_i}s_j(r_b-\alpha m_j)_\theta,
  \end{equation}
  which depends not only on the value of $s_i$, but on a
subset of $S$, namely $\bigcup_{j\in N_i} P_j$, which
combines the two interaction neighborhoods, $\{P_j\}$
controlling birth-death competition within a range $R$, and
$N_i$ implementing excluded volume effects. \corr{Note that
this contribution is a natural extension to the one
proposed in \cite{helo}, but adapted to an
occupation--limited \ehg{dynamics}, that is, now $\pi_b$
can vanish because \ehg{of $s_i=\sigma_i$ }(node is full)
without being $r_b\le \alpha m_j$}.
\item Movements: A particle at node $i$ moves to node $j\in
    N_i$ with rate $r_m(\sigma_j-s_j)$, i.e. jumps are more
    likely to occur to emptier sites. The total rate of
    movements from node $i$ to $j\in N_i$ is
  \begin{equation}
    \label{eq:3}
    \pi_m(s_i,s_j)=r_ms_i(\sigma_j-s_j).
  \end{equation}
\end{itemize}

\begin{figure}[!h]
  \centering
  \includegraphics[width=.75\textwidth]{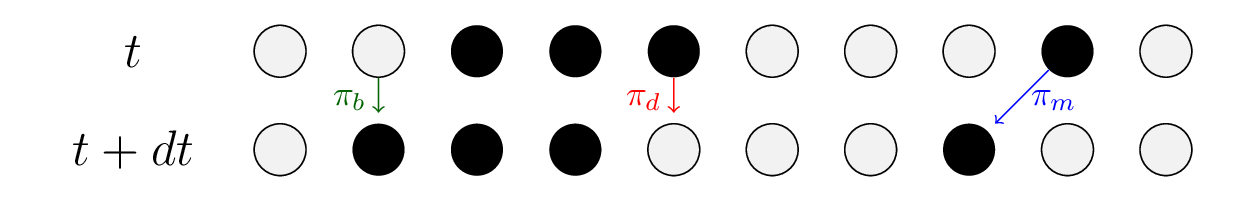}
  \caption{Schematic representation of the time evolution of the system in 1D. \ehg{Assuming $\sigma_i=1$,
  grey circles represent empty nodes and black circles occupied ones.} }
  \label{fig:01}
\end{figure}

\corr{In figure \ref{fig:01} we schematically represent the
system evolution for the one dimensional case} and in Table
\ref{table} we summarize the parameters of the model.

\begin{table}[!h]
  \centering
  \begin{tabular}[!h]{r|l}
    Parameter & Meaning \\ \hline
    $\Sigma$ (Greek S) & Hypercubic lattice of dimension $d$ \\
    $s_i$ & Number of particles at node $i$   \\
    $\sigma_i$ & Maximum value of $s_i$  \\
    $N_i$ (Greek N) & $\{i\}\cup \{j\in \Sigma \ : \ j$ is a first neighbor of $i\}$ \\
    $n_i$ & Number of particles in $N_i$ \\
    $\nu_i$ & Maximum value of $n_i$  \\
    $P_i$ (Greek R) & $\{j\in \Sigma \ :\ |\vec r(i)-\vec r(j)|\le R|\}$  \\
    $V_R$ &  $2 \pi^{d/2} R^d /[d\Gamma(d/2)]$,  the $d$-volume of $P_i$ \\
    $m_i$ & Number of particles in $P_i$  \\
    $r_d$ & Rate of death of a particle  \\
    $\alpha$ & Positive competition parameter \\
    $r_b$ & Competition-independent part of the particle rate of birth per accessible site \\
    $r_m$ & Particle rate of jumps per accessible site from node $i$ to $j\in N_i$\\
    $c_1$ & $2 d r_b/r_d$  \\
    $c_2$ & $2d\alpha \rho_m V_R/r_d$ \\
    $c_3$ &  $dr_m/(N^{2/d}r_d)$  \\
    $c_4$ & $r_m/[2\alpha (N R/L)^3] $, parameter used in the 1d case  \\

  \end{tabular}
  \caption{Some parameters and constants used along the work}
  \label{table}
\end{table}

\subsection{Moment equations}

From the latter rates, the master equation for the probability
$p(S,t)$ of the system being at state $S$ at time $t$ can be
written as
\begin{eqnarray}
  \label{eq:4}
  \frac{\partial}{\partial t}p(S,t)=\sum_{i\in \Sigma}
  & \biggl\{
  (E_i^+-1)\left[\pi_d(s_i)p(S,t)\right]+(E_i^--1)\left[\pi_b(s_i,S)p(S,t)\right]  \nonumber  \\
  &   +\sum_{j\in N_i-\{i\}}(E_i^+E_j^--1)\left[\pi_m(s_i,s_j)p(S,t)\right]
  \biggr\},
\end{eqnarray}
 where we have introduced the following operators \cite{ka92}:
 \begin{eqnarray}
   \label{eq:5}
   &E_i^+f(s_1,\dots,s_i,\dots,s_N)=f(s_1,\dots,s_i+1,\dots,s_N), \\
   \label{eq:6}
   &E_i^-f(s_1,\dots,s_i,\dots,s_N)=f(s_1,\dots,s_i-1,\dots,s_N).
 \end{eqnarray}
The master equation must be solved with any initial normalized and positive $p(S,0)$ verifying
\begin{equation}
  \label{eq:7}
  p(S,t)=0, \qquad \textrm{if} \quad s_i<0 \quad \textrm{or} \quad s_i>\sigma_i,
\end{equation}
for any $i\in\Sigma$ and $t=0$. Equation \eqref{eq:4}
guarantees that $p$ is normalized, positive, and condition
\eqref{eq:7} fulfilled for any $t\ge 0$.

By taking moments of the master equation \eqref{eq:4} and using
condition \eqref{eq:7}, we obtain the following equations for
the first moments of $p(S,t)$,
\begin{eqnarray}
  \label{eq:8}
  \frac{d}{dt} & \medio{s_i}=-r_d\medio{s_i}+\sum_{j\in N_i}\medio{(r_b-\alpha m_j)_\theta s_j(\sigma_i-s_i)} \\
  & \nonumber  \qquad  +r_m(\sigma_i \medio{n_i}-\nu_i\medio{s_i}), \\
  \label{eq:9}
  \nonumber
  \frac{d}{dt} &\medio{s_i^2} = \medio{\pi_d(s_i)+\pi_b(s_i,S)} +\medio{2s_i\left[\pi_b(s_i,S)-\pi_d(s_i)\right]}\\
  & \qquad +r_m\sum_{j\in N_i-\{i\}}\medio{\sigma_is_j(1+2s_i)+\sigma_js_i(1-2s_i)-2s_is_j}, \\
  \label{eq:10}
  \nonumber
  \frac{d}{dt} &\medio{s_is_j} =-2r_d\medio{s_is_j} +\medio{\pi_b(s_i,S)s_j+\pi_b(s_j,S)s_i} \\
  & \nonumber  \qquad +r_m\left[\sigma_i\medio{n_is_j}+\sigma_j\medio{s_in_j}-(\nu_i+\nu_j)\medio{s_is_j}\right. \\
  & \qquad \qquad \left. -\left(\sigma_i\medio{s_j}+\sigma_j\medio{s_i} -2\medio{s_is_j}\right)\delta_{i\in N_j-\{j\}}\right], \ j\ne i
\end{eqnarray}
for $i,j\in \Sigma$. We have introduced two new quantities,
namely the total number of particles in $N_i$:
\begin{equation}
  \label{eq:11}
  n_i\equiv \sum_{j\in N_i}s_j, \\
\end{equation}
and its maximum possible value:
\begin{equation}
  \label{eq:12}
  \nu_i\equiv \sum_{j\in N_i}\sigma_j.
\end{equation}
The function $\delta_{i\in A}$ returns $1$ if $i$ belongs to a
set $A$ and $0$ otherwise. \corr{Each of the three equations
\eqref{eq:8}--\eqref{eq:10} contain three terms on their r.h.s.
that account the three processes introduced in the model. If we
consider Eq. \eqref{eq:8} as an example, it is apparent that
the term proportional \ehg{to $r_d$ (death term) decreases the
mean number of particles of node $i$. The second term accounts
for the births, with a rate dependent on the occupation of the
nearest neighbor sites and of the competition neighborhood
$P_i$. Finally, the third contribution reflects particle
motion.}  }

The system of equations \eqref{eq:8}--\eqref{eq:10}, for
$i,j\in\Sigma$, is exact but not closed, due to the presence of
moments of higher orders. In general, this problem can be
circumvented by making the following sequence of
approximations:
First, we factorize correlations involving $\pi_b$:
\begin{equation}
  \label{eq:13}
\nonumber  \medio{\pi_b(s_i,S) f(S)} \simeq \sum_{j\in N_i} \medio{(r_b-\alpha m_j)_\theta} \medio{s_{j}(\sigma_i-s_i)f(S)} ,
\end{equation}
for any function $f(S)$. Second, we also approximate
\begin{equation}
  \label{eq:14}
  \medio{(r_b-\alpha m_{i})_\theta}\simeq (r_b-\alpha \medio{m_{i}})_\theta.
\end{equation}
These two approximations are expected to be correct if
$(r_b-\alpha m_{i})_\theta$ has small fluctuations, which could
be the case when $m_{i}$ involves a large number of particles,
i.e., when $R$ is big enough, and the mean density of particles
is also large. Third, three-node correlations are neglected as
\begin{equation}
  \label{eq:15}
  \medio{\left(s_i- \medio{s_i}\right)\left(s_j- \medio{s_j}\right)\left(s_k- \medio{s_k}\right)}\simeq 0,
\end{equation}
which allows us to express $\medio{s_is_js_k}$ as sums of the
form $\medio{s_i s_j}\medio{s_k}$. This is a good approximation
since the main source of correlations in the model is via
particles births, that take place among neighbor node pairs.

It is worth to mention that the latter simplification implies
some time limitations to the resulting equations. In
particular, if $r_d\ne 0$, there is a positive probability for
the stochastic particle system to become extinct for any given
values of the rest of parameters, implying that $\medio{s_i}$
always decays to zero if we wait long enough. In contradiction
to that, we anticipate that the simplified equations have
stable steady state solutions with $\medio{s_i}\ne 0$.
Nevertheless we expect, and corroborate with numerical
simulations, that the provided simplified description is
accurate in a wide time window, useful for understanding
relevant observable processes.

With approximations \eqref{eq:13}--\eqref{eq:15}, Eqs.
\eqref{eq:8}--\eqref{eq:10} become a \ehg{closed system, but
difficult to deal with because of the large number ($N+N^2$) of
coupled equations}. The description can be simplified if we
pass to the continuum limit. Two cases will be considered in
the next section, namely when there is no restriction on the
number of particles at a node (no volume exclusion or
$\sigma_i\gg 1$) which is the case previously studied in
\cite{helo}, and the novel case of (extreme) volume exclusion
where at most one particle can be present in one node.

\section{Macroscopic description}\label{sec:macro}

\subsection{Macroscopic description without volume exclusion}

Suppose each node can have a large number of particles,
$\sigma_i= \sigma \gg 1$ for all $i\in\Sigma$. Then, except for
extreme densities, any term of the form $\sigma_i-s_j$ reduces
to $\sigma$. As a consequence, Eqs. in \eqref{eq:8} with
approximations \eqref{eq:13} and \eqref{eq:14}, regardless of
approximation \eqref{eq:15}, decouples from the rest, leading
to
\begin{equation}
  \label{eq:16}
  \frac{d}{dt} \medio{s_i}=-r_d\medio{s_i}+\sigma \sum_{j\in N_i}(r_b-\alpha \medio{m_j})_\theta \medio{s_j}
  +r_m\sigma \left[\medio{n_i}-(2d+1)\medio{s_i}\right],\
\end{equation}
for $i=1,\dots, N$. Since $\sigma_i=\sigma$, we have used the fact that $\nu_i=(2d+1)\sigma$.

We now pass to the continuum limit in which the number of nodes
$N$ in a system of fixed size $L$ increases,
$N\rightarrow\infty$, or equivalently the lattice spacing
vanishes $\Delta\equiv L/N^{1/d}\rightarrow 0$. We use the
vector position $\vec r(i)$ instead of label $i$, $P_{\vec r}$
instead of $P_i$ (with the same meaning), and the expected
value of the density of particles $\rho(\vec r)$ instead of
$\medio{s_i}$, with the following identifications
\begin{eqnarray}
  \label{eq:17}
   & i \to \frac{\vec r}{\Delta} ,  \\
 \label{eq:18}
   & \medio{s_i} \to \frac{\rho(\vec r,t)}{\rho_m} ,
\end{eqnarray}
where
\begin{equation}
  \label{eq:19}
  \rho_m\equiv \frac{N}{L^d} = \Delta^{-d}
\end{equation}
is the spatial density of lattice nodes. $\sigma\rho_m$ is the
maximum value allowed to the particle density. For $j\in N_i$
we have $|\vec r(i)-\vec r(j)|=\Delta$, which is much smaller
than $L$ for $N\gg 1$. Hence, the continuum limit is a good
approximation if $N\gg 1$ for fixed $L$ and $\rho(\vec r,t)$ is
a smooth function of $\vec r$, with the following results
\begin{eqnarray}
  \label{eq:20}
  (r_b-\alpha \medio{m_j})_\theta  & \rightarrow &
  \left(r_b-\alpha \int_{P_{\vec r}} \rho(\vec r',t) d\vec r'\right)_\theta+\mathcal{O}\left(N^{-1}\right),\\
  \label{eq:21}
  \left[\medio{n_i}-(2d+1) \medio{s_i}\right] & \rightarrow &
  \frac{L^2}{N^{\frac{2}{d}}}\frac{1}{\rho_m}\nabla^2 \rho +\mathcal{O}\left[N^{-\frac{2}{d}-1}\right], \\
  \label{eq:22}
\nonumber
   \sum_{j\in N_i} (r_b-\alpha \medio{m_j})_\theta \medio{s_j} & \rightarrow &
  2d \left(r_b-\alpha \int_{P_{\vec r}}
 \rho(\vec r',t) d\vec r'\right)_\theta  \frac{\rho}{\rho_m} \\
 &&  \qquad + \mathcal{O}\left[N^{-\frac{2}{d}}(\nabla \rho)^2, N^{-\frac{2}{d}}\nabla^2 \rho \right].
\end{eqnarray}
In some limits, for example $r_m=0$, the omitted terms in the
last expression should be kept in order to account for residual
diffusion processes. Finally we arrive at
\begin{eqnarray}
  \label{eq:23}
  \frac{\partial \rho(\vec r,t)}{\partial t}=&-r_d\rho+\sigma \frac{r_m}{\rho_m^{2/d}}\nabla^2 \rho +
  2d\sigma \left[r_b-\alpha \int_{P_{\vec r}}\rho(\vec r~',t)d\vec r~'\right]_\theta \rho,
\end{eqnarray}
which is, with an appropriate redefinition of constants, the
same non-local Fisher-Kolmogorov equation derived and studied
in previous works
\cite{helo,lohe04,lohe07,hehelo15,Fuentes2003}.
In these papers it was justified that the maximum condition for the
Heaviside function term was rarely needed so that
$\left[r_b-\alpha \int_{P_{\vec r}}\rho(\vec r~',t)d\vec r~'\right]_\theta
\approx (r_b-\alpha \int_{P_{\vec r}}\rho(\vec r~',t)d\vec r~')$
 Note that,
since $\rho_m^{2/d}=\Delta^{-2}$, we recover the usual
situation in which a large value of the total jumping rate
$\sigma r_m$ is needed to keep a nonvanishing diffusion
coefficient $\sigma r_m \Delta^2$ in the continuum limit
$N\to\infty$ or $\Delta\to 0$.

\subsection{Macroscopic description with volume exclusion}

We consider now the situation of full volume exclusion, i.e. at
a given time at most one particle can be at a node: $\sigma_i=
1~\forall i\in\Sigma$. In this case the lattice node density
$\rho_m$ becomes also the maximum possible particle density. We
have $\nu_i=2d+1$, and $s_i(1-s_i)=0$, or $s_i^2=s_i$, so that
Eq. \eqref{eq:9} becomes irrelevant. We still have to deal with
the system of equations \eqref{eq:8} and \eqref{eq:10}.

If we take into account approximations
\eqref{eq:13}--\eqref{eq:15}, the equation \eqref{eq:8} for
$\medio{s_i}$ does not involve terms of the form
$\medio{s_is_j}$ for $j\notin N_i$.
As a consequence, in passing to the continuum limit, all we need are
the identifications in Eqs. \eqref{eq:17} and \eqref{eq:18},
and one more for terms like $\medio{s_is_j}$ for $j\in N_i$. In
a statistically isotropic system the needed identification
should have the form
\begin{equation}
  \label{eq:24}
  \medio{s_is_j} \rightarrow \frac{1}{\rho_m^2} \kappa(\vec r,t),
  \qquad \textrm{for any } j\in N_i, \textrm{and } i\to \vec r/\Delta \ ,
\end{equation}
where $\kappa$ is a scalar, rather than a tensor, function.
Note that $\kappa$ takes into account the short-range
(nearest-neighbor) correlations in the system. The continuum
equations, to leading order in $1/N$, reduce to
\begin{eqnarray}
  \label{eq:25}
  \nonumber
  & \frac{\partial}{\partial t}\rho(\vec r,t) =-r_d\rho +\frac{r_m}{\rho_m^{2/d}} \nabla^2
\rho+  \\
& \qquad 2d \left[r_b-\alpha \int_{P_{\vec r}}\rho(\vec r~',t)d\vec r~'\right]_\theta \left(\rho-\frac{\kappa}{\rho_m}\right),\\
  \label{eq:26}
  \nonumber & \frac{\partial}{\partial t}\kappa(\vec r,t) = -2r_d\kappa -2dr_m\left(\kappa-\rho^2\right)  \\
  & \qquad + 2d \left[r_b-\alpha \int_{P_{\vec r}}\rho(\vec r~',t)d\vec r~'\right]_\theta \rho_m\rho\left[1+2\left(\frac{\rho}{\rho_m}\right)^2-3\frac{\kappa}{\rho_m^2}\right].
\end{eqnarray}

System \eqref{eq:25}--\eqref{eq:26} can be reduced to a single
equation for the density in two interesting cases that we
consider in the following.

\subsubsection{Correlation factorization at intermediate times.}
\label{subsubsec:factor}

Suppose first that there are no deaths nor births in the
system, $r_d=r_b=\alpha=0$, so that only the terms containing
the jump rate $r_m$ remain in Eqs. \eqref{eq:25}-\eqref{eq:26}.
Eq. \eqref{eq:26} implies that $\kappa$ approaches $\rho^2$ in
a time of the order of $1/(2dr_m)$, whereas relaxation times of
the density, according to Eq. \eqref{eq:25}, are of the order
of $(l/\Delta)^2/r_m$, with $l$ the typical length of variation
of $\rho$. In the continuum description this time is long since
$l/\Delta \gg 1$, so that for time scales larger than
$1/(2dr_m)$ we can approximate the time-dependence of $\kappa$
via the density as:
\begin{equation}
  \label{eq:27}
  \kappa(\vec r,t) \simeq \left[\rho(\vec r,t)\right]^2.
\end{equation}
For general values of the rates $r_d$, $r_b$ and $\alpha$, Eq.
\eqref{eq:27} will be still approximately valid for times
larger than $1/(2dr_m)$ provided that $r_m$ is large enough so
that the corresponding term in Eq. \eqref{eq:26}, describing
particle motion, dominates the others. Regardless the relative
value of the rates, but if $\rho\simeq \rho_m$, the latter
approximation also holds, since now almost all nodes are
filled. Using Eq. \eqref{eq:27} together with Eq.
\eqref{eq:25}, we get

\begin{equation}
  \label{eq:28}
\frac{\partial}{\partial t}\rho(\vec r,t) = -r_d\rho +\frac{r_m}{\rho_m^{2/d}} \nabla^2 \rho
+ 2d \left[r_b-\alpha \int_{P_{\vec r}}\rho(\vec r~',t)d\vec r~'\right]_\theta \left(1-\frac{\rho}{\rho_m}\right)\rho. \
\end{equation}
This equation can also be derived directly by taking the
continuum limit of Eqs. \eqref{eq:8} and \eqref{eq:10},
together with approximations \eqref{eq:13}--\eqref{eq:14}, and
factorizing correlations among all nodes, namely
$\medio{s_is_j}\simeq \medio{s_i}\medio{s_j}$ for all $i,j\in
\Sigma$ and $i\ne j$.

Equation \eqref{eq:28} is the central result of this paper and
will be analyzed in detail in the following sections. Two main
comments arise about this expression: i) The diffusion constant
appearing as the coefficient of the Laplacian term is
$r_m/\rho_m^{2/d}=r_m\Delta^2$, independent of particle
density, as in cases of non-reacting particles with full volume
exclusion ($\sigma=1$) \cite{chikrma14}. We do not expect this
simplicity to remain beyond the $\sigma=1$ and $\sigma=\infty$
limits. ii) The effect of volume exclusion is in the specific
form of the birth term (the last one in the r.h.s). Comparing
to the no-exclusion case \eqref{eq:23} the difference is the
additional factor $(1-\rho/\rho_m)$. It always reduces the
birth term and gives a vanishing birth rate when all nodes are
occupied. We note that Eq. \eqref{eq:28} is different from
other non-local cubic equations in the literature
\cite{Clerc2010,Escaff2015}. If the range of competitive
interaction is much smaller than the typical length-scale
variation of $\rho$ (i.e. $R\to 0$), \corr{which is the case,
\ehg{for example,} when $\rho$ is nearly homogeneous in space},
the birth term reduces to $2d(r_b-2\alpha
V_R\rho)(1-\rho/\rho_m)\rho$ and contains a term proportional
to $\rho^3$ (where $V_R=\int_{P_{\vec r}} d\vec r' = \frac{2
\pi^{d/2}}{d\Gamma (d/2)}R^d$ is the volume of the
$d$-dimensional sphere of radius $R$).

\subsubsection{The case of small density.}
\label{subsubsec:dilute}

Another limit in which \eqref{eq:25}--\eqref{eq:26} reduces to
a single equation for the density corresponds to $\rho\ll
\rho_m$, i.e. a very dilute system. In this case, it is
convenient to expand $\kappa$ in powers of $\rho$, so that at
leading order $\kappa \propto \rho$. Keeping only terms linear
in the density and taking into account that for $\rho$ small
the argument of the $\theta$ function is positive, Eq.
\eqref{eq:26} reduces to
\begin{equation}
  \label{eq:29}
  \frac{\partial}{\partial t}\kappa(\vec r,t) = -2r_d\kappa -2dr_m\kappa+ 2d r_b \rho_m\rho.
\end{equation}
Once again, when comparing the time-scales of evolution of
$\kappa$ in Eq. \eqref{eq:29} with the ones for $\rho$ in Eq.
\eqref{eq:25} we conclude that, when the jumping rate $r_m$
dominates the other rates, $\kappa$ is related to $\rho$ via
the steady--state expression obtained from \eqref{eq:29}:
\begin{equation}
  \label{eq:30}
  \kappa (\vec r,t)\simeq \frac{dr_b}{r_d+dr_m}\rho_m \rho(\vec r,t).
\end{equation}
Within this approximation, we get the following equation for the density,
\begin{eqnarray}
  \label{eq:31}
\nonumber \frac{\partial}{\partial t}\rho(\vec r,t) =&-r_d\rho +\frac{dr_m}{\rho_m^{2/d}} \nabla^2 \rho  \\
& + 2d \left(1-\frac{d r_b}{r_d+dr_m}\right) \left[r_b-\alpha \int_{P_{\vec r}}\rho(\vec r~',t)d\vec r~'\right]_\theta  \rho(\vec r,t).
\end{eqnarray}
This equation has the same structure as Eq. \eqref{eq:23},
obtained without volume exclusion, but with different
parameters. Here also volume exclusion reduces the effective
birth rate. For consistency, since the birth term should be
positive, Eq. \eqref{eq:31} can only apply if $r_b\le (r_m +
r_d/d)$.

\section{Homogeneous solutions and spatial patterns for the excluded-volume birth-death equation}
\label{sec:4}

In this section we analyze Eq. \eqref{eq:28}. As commented
before, this is the extension, to include volume exclusion
effects, of the already studied spatially non-local birth-death
competition model in
\cite{helo,lohe04,lohe07,hehelo15,Fuentes2003}. This last
equation supports homogeneous solutions as well as spatially
periodic ones. We analyze here how volume exclusion modifies
them. First, it is worth writing Eq. \eqref{eq:28} as
\begin{eqnarray}
  \label{eq:32}
  \nonumber
  \frac{\partial}{\partial s}\rho(\vec r,s) &=& -\rho + \left[c_1-c_2\frac{1}{V_R\rho_m}
 \int_{P_{\vec r}}\rho(\vec r',s)d\vec r'\right]_\theta \left(1-\frac{\rho}{\rho_m}\right)\rho \\ && +c_3L^2 \nabla^2 \rho,
\end{eqnarray}
where we have defined a new time scale $s=r_d t$, and
introduced new non--dimensional constants
\begin{equation}
  \label{eq:33}
  c_1\equiv \frac{2dr_b}{r_d}, \qquad c_2\equiv \frac{2d\alpha \rho_m V_R}{r_d},
  \qquad c_3\equiv \frac{dr_m}{N^{2/d}r_d}=\frac{d r_m}{r_d}(\Delta/L)^2 .
\end{equation}
See a summary of the notation in Table \ref{table}. Each
constant is proportional to a specific rate measured in units
of the death rate: $c_1$ is proportional to the birth rate and
$c_3$ to the rate of movements. $c_2$ is proportional to the
rate associated to resources competition $\alpha$ and we have
also introduced in it the quantity $\rho_m V_R$ which is the
maximum number of particles in a competition neighborhood
$P_{\vec{r}}$. This makes more explicit that, if we express all
length scales in units of $L$ or of $R$ and write Eq.
\eqref{eq:33} in terms of $\rho/\rho_m \in [0,1]$, the only
relevant parameter additional to $c_1$, $c_2$ and $c_3$ is
$R/L$. In a large system ($L\to\infty$) the relevant parameters
are just $c_1$, $c_2$ and $c_3$. This contrasts with  the
case without volume exclusion, Eq. \eqref{eq:23}, which can be
written in terms of a single dimensionless parameter for large
$L$ \cite{lohe04}.

\subsection{Homogeneous evolution and steady states}
If we focus on spatially homogeneous time-dependent solutions
$\rho(x,s)=\rho_h(s)$, Eq. \eqref{eq:32} reduces to
\begin{equation}
  \label{eq:34}
  \frac{d \rho_h(s)}{d s}=-\rho_h(s)+\left[c_1-c_2\frac{\rho_h(s)}{\rho_m}\right]_\theta
     \left[1-\frac{\rho_h(s)}{\rho_m}\right]\rho_h(s).
\end{equation}
Suppose initially $\rho_h(0)\ge \rho_m \frac{c_1}{c_2}$, then
the dynamics evolves in two stages:
\begin{itemize}
\item [(i)] For $s\in[0,s_0]$ with $s_0\equiv\ln
    \frac{\rho_h(0) c_2}{\rho_m c_1}$, we have (note that
    the term containing the Heaviside function vanishes)
  \begin{equation}
    \label{eq:35}
    \rho_h(s)=\rho_h(0)e^{-s}.
  \end{equation}
\item [(ii)] For $s\ge s_0$,  $\rho_h(s)\le \rho_m
    \frac{c_1}{c_2}$ and the dynamics becomes more
    involved. But, since Eq. \eqref{eq:34} is  \corr{an autonomous} first-order
    ordinary differential equation, the only possible
    behavior is a monotonous approach to one of the
    available fixed points.
\end{itemize}
If initially $\rho_h(0) \le \rho_m \frac{c_1}{c_2}$, the
dynamics is always at stage (ii). Hence, the system always
approaches the steady states within stage (ii).

We can obtain explicit expressions for these steady--state
homogeneous solutions $\rho_h(\infty)$. They are obtained by
equating to zero the r.h.s of Eq. \eqref{eq:34}, with the
following result,
\begin{equation}
  \label{eq:36}
  \rho_h(\infty)=\left\{
    \begin{array}{lcr}
      \rho_0=0 &\textrm{ for } & c_1,c_2\ge 0, \\
      \rho_1=\rho_m \left(\frac{c_1+c_2}{2c_2}-\frac{1}{2c_2}\sqrt{(c_1-c_2)^2+4c_2} \right) &
                        \textrm{ for } & c_1\ge 1,\ c_2\ge 0.
    \end{array}
 \right.
\end{equation}
For the parameters of the system that make $c_1\le 1$ (we
recall that $c_2\ge 0$ in this work, since competitive
interactions require $\alpha>0$) only the $\rho_0$ solution
exists, representing complete extinction of the population,
while it coexists with another homogeneous solution for $c_1>
1$. In this last case also a third steady homogeneous solution
appears, but it is non-physical since it would give $\rho\ge
\rho_m$.

\subsection{Linear stability analysis of the homogeneous solutions and pattern formation}

The stability of the solutions in Eq. \eqref{eq:36} can be
determined by applying standard linear stability analysis. By
adding a perturbation of small amplitude $A$ to the homogeneous
states we obtain the linear evolution as
\begin{equation}
  \label{eq:37}
  \rho(\vec r,s)= \rho_h(\infty)+A\exp\left(\lambda_{\vec n} s+i\vec{q}_{\vec n}\cdot \vec r\right),
\end{equation}
where $\vec{q}_{\vec n}\equiv \frac{2\pi}{L}\vec n$, in terms
of the $d$-dimensional vectors of integers $\vec n$, are the
wavevectors allowed by the periodic boundary conditions.

The specific values of the growth rates $\lambda_{\vec n}$
depend on the steady--state solution considered, $\rho_0$ or
$\rho_1$. For the extinction solution, $\rho_0=0$, we have
  \begin{equation}
    \label{eq:38}
    \lambda_{\vec n}=-\left[1-c_1+c_3(2\pi \vec n)^2\right] .
  \end{equation}
Since $\lambda_{\vec n}\le \lambda_{\vec 0}=-(1-c_1)$, the
solution is linearly stable for $c_1< 1$ and unstable for $c_1
> 1$. In this last case the fastest growth rate occurs for
$\vec n=\vec 0$. Therefore we expect the homogeneous solution
$\rho_1$ to develop after this instability.

When linearizing close to $\rho_h(\infty)=\rho_1$, the growth
rate is
\begin{equation}
  \label{eq:39}
  \lambda_{\vec n}=-\left[\frac{\rho_1}{\rho_m-\rho_1}+c_3(2\pi \vec n)^2
                    +c_2\frac{\rho_1 (\rho_m-\rho_1)}{2\rho_m^2} \mathcal{F}(\vec q_{\vec n})\right],
\end{equation}
where $\mathcal{F}(\vec q_{\vec n})$ is the Fourier transform
of the Heaviside function $\theta(R-|\vec r|)$. For large $c_3$
(large diffusion) or $c_2$ small enough (weak competition or
small $R$), and since $\mathcal{F}(\vec q_{\vec n})$ is a
bounded function, $\lambda_n$ is negative for all possible
$\vec n$. On the contrary, if $c_2$ becomes large enough, there
are values of $\vec n$ for which $\lambda_{\vec n}$ becomes
positive. We expect then that the density will develop periodic
patterns with a periodicity given approximately by the
wavenumber $\vec q_{\vec n}$ which first becomes unstable. From
generic arguments considering the competition dynamics
\cite{hehelo15}, this periodicity would be in between $R$ and
$2R$. The onset of instability is located by solving the
following problem
\begin{equation}
  \label{eq:40}
  \lambda_{\vec n}=0, \qquad \vec \nabla_{\vec n} \lambda_{\vec n}=\vec 0.
\end{equation}
Here we solve it explicitly in the one-dimensional case. Now
the vectors $\vec n$ and $\vec q_{\vec n}$ become scalars, $n$
and $q_n=2\pi n/L$, and the Fourier transform
$\mathcal{F}(q_n)$ is
\begin{equation}
  \label{eq:41}
  \mathcal{F}(q_n)=2\frac{\sin(\gamma_n)}{\gamma_n},
  \qquad \gamma_n \equiv q_n R = \frac{2\pi R}{L}n, \qquad n=0,\pm 1, \pm 2, ...
\end{equation}
After some algebra, problem \eqref{eq:40} for the onset of
instability reduces to
\begin{eqnarray}
  \label{eq:42}
  &c_4\left(\gamma_n\right)^2 =\frac{\rho_1}{\rho_m}\left(1-\frac{\rho_1}{\rho_m}\right)\left[\frac{\sin\left(\gamma_n\right)}{\gamma_n}-\cos\left(\gamma_n\right)\right], \\
  \label{eq:43}
  &2 =c_2\left[1-\frac{\rho_1}{\rho_m}\right]^2\left[\cos\left(\gamma_n\right)-3\frac{\sin\left(\gamma_n\right)}{\gamma_n}\right],
\end{eqnarray}
where
\begin{equation}
  \label{eq:44}
  c_4\equiv \frac{r_m}{2\alpha}\frac{1}{\left(\rho_m R\right)^3}
\end{equation}
is a new parameter that compares the importance of movement
with respect to competition. $\rho_m R$ is the maximum number
of particles in a competition region of size $R$. We recall
that the steady-state solution $\rho_1/\rho_m$ depends also on
the constants $c_1$ and $c_2$ as Eq. \eqref{eq:36} shows, hence
Eqs. \eqref{eq:42} and \eqref{eq:43} depend on the parameters
of the system through $c_1$, $c_2$ and $c_4$. We can also look
at Eqs. \eqref{eq:42}--\eqref{eq:43} as defining a surface of
critical points in the space of parameters $(c_1,c_2,c_4)$,
with a critical value of $\gamma_n$ assigned to each of these
points. From the value of $\gamma_n$ one can get the wavenumber
$q_n$ or $n$, which determine the expected periodicity of the
pattern-forming at each instability point as $l\approx
2\pi/q_n$. As expected \corr{from the explanation \ehg{of the
instability mechanism} provided in \cite{hehelo15}} this
periodicity is in between $R$ and $2R$.

\begin{figure}[!h]
  \centering
  \includegraphics[width=.7\textwidth]{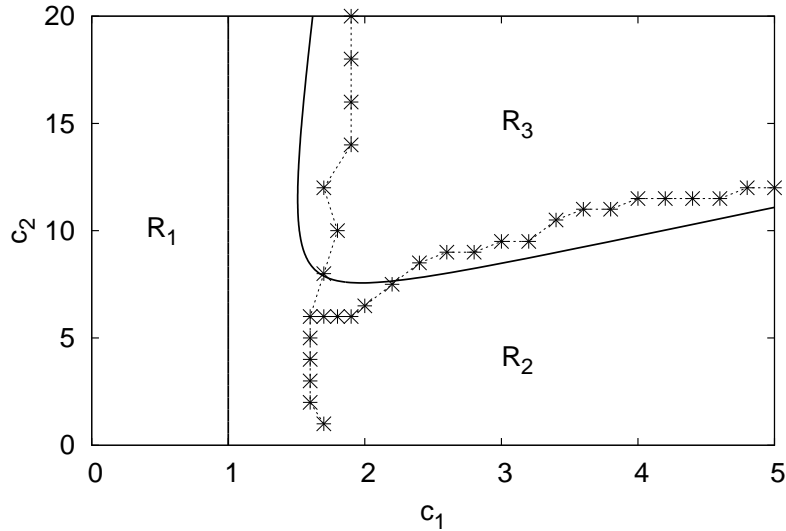}
  \caption{Phase diagram indicating three different stability regions (separated by the black lines) from the linear analysis of
  homogeneous solutions of Eq. \eqref{eq:28}: in R$_1$ the vanishing homogeneous state, $\rho_0=0$,
  is stable. In R$_2$ the non-vanishing homogeneous state $\rho_1$ is stable. In  R$_3$ the homogeneous solutions
  are unstable and periodic patterns would appear. Coefficients $c_1$ and $c_2$ are defined in \eqref{eq:33} while $c_4$ is given by \eqref{eq:44} and is  $c_4=5\cdot 10^{-4}$ for the present case. Line with asterisks
  indicates the transitions obtained from particle simulations where, besides the already
  stated parameters, $N=2240$ and $R/L=0.1$.}
  \label{fig:phase}
\end{figure}

The numerical evaluation of Eqs. \eqref{eq:42} and
\eqref{eq:43} for a particular value of $c_4$ is shown in Fig.
\ref{fig:phase} (solid lines). The different regions that we
have unveiled, that is, stable zero solution, stable non-zero
homogeneous solution, and pattern forming region (corresponding
to the instability of $\rho_1$) are labelled as R$_1$, R$_2$
and R$_3$, respectively. Note that the transition from the zero
solution to the spatial pattern, occurring for a sufficiently
large fixed value of $c_2$ by increasing $c_1$, necessarily
passes through the non-zero homogeneous solution (region
R$_2$). But this is not the case numerically observed by
simulating the stochastic model (asterisks in the figure) as is
discussed in the next section.

\ehg{It is worth mentioning that the general formulae of our
stability analysis (Eqs. (\ref{eq:38}) and (\ref{eq:39})) are
valid in arbitrary dimension and set the qualitative behavior
beyond the one-dimensional case on which we focus in this
paper. Namely, the phase diagram involves three independent
parameters, and the system may exhibit three distinct phases:
extinction, spatially homogeneous, and periodic patterns (with
dimension-dependent shapes).}

\section{Numerical simulations of the particle dynamics. Comparison with macroscopic description
\label{sec:5}}

In this section we compare the theoretical results of Sec.
\ref{sec:4} with numerical simulations, in one dimension and
with full volume exclusion ($\sigma=1$), of the stochastic
process defined by the rates given in Eqs.
\eqref{eq:1}-\eqref{eq:3}. For the parameter values we use
$N=2240$, $R/L=0.1$, and $r_d=L=1$ (this fixes the units of
time and of space, respectively) 
The quantities $r_b$, $\alpha$
and $r_m$ (or equivalently $c_1$, $c_2$ and $c_3$, and hence
$c_4$) have been varied.
%
%
In the sequel, we explain the algorithm of simulations and then
focus on the comparison itself.

\subsection{Numerical simulations}
We use  the Gillespie  algorithm \cite{gi77,hehelo12} to
simulate a regular one dimensional system with periodic
boundary conditions. An arbitrary initial configuration $S$ of
the system is updated in several steps:
\begin{itemize}
\item [(i)] The rates at each node $i=1,\dots N$,
    associated to the three processes $\pi_d(s_i)$,
    $\pi_b(s_i,S)$ or
    $\left[\pi_m(s_i,s_{i-1})+\pi_m(s_i,s_{i+1})\right]$
    are calculated,  as well as the total rates, given by
\begin{eqnarray}
  \label{eq:45}
  & \Pi_d=\sum_{i=1}^N \pi_d(s_i), \\
  \label{eq:46}
  & \Pi_b=\sum_{i=1}^N \pi_b(s_i,S), \\
  \label{eq:47}
  & \Pi_m=\sum_{i=1}^N\left[\pi_m(s_i,s_{i-1})+\pi_m(s_i,s_{i+1})\right].
\end{eqnarray}
 The current time $t$ is incremented as $t\to t+ \Delta t$,
 where $\Delta t$ is generated with the following
 exponential density probability
\begin{equation}
  \label{eq:48}
 p(\Delta t)=(\Pi_d+\Pi_b+\Pi_m)^{-1}\exp\left[(\Pi_d+\Pi_b+\Pi_m) \Delta t \right].
\end{equation}
\item [(ii)] After this time $\Delta t$, one of the three
    possible events (death, birth or movement) is selected
    with probability proportional to their corresponding
    rates ($\Pi_d$, $\Pi_b$ and $\Pi_m$).
\item [(iii)] One node, say $i$, is selected with
    probability proportional to its contribution to the
    rate of the selected process ($\pi_d(s_i)$,
    $\pi_b(s_i,S)$ or
    $\left[\pi_m(s_i,s_{i-1})+\pi_m(s_i,s_{i+1})\right]$).
\item [(iv)] Depending on the specific process selected in
    (ii) and the node involved (iii), the state $S$ of the
    system is updated: $s_i\to 0$ (death), $s_i\to 1$
    (birth), and $s_i\to 0, \ s_j\to 1$ (movement), where
    $j$ is selected equiprobably among the empty next
    neighbors of $i$.
\item [(v)] If $S$ corresponds to an empty configuration or
    the desired total time of simulation has been reached,
    the simulations finishes. In other cases we go back to
    (i) with the updated state. At some intermediate steps,
    and under some conditions, properties of interest are
    measured from the simulations.
\end{itemize}

\subsection{Small birth rates. Decay to the extinction solution}

\begin{figure}[!h]
  \centering
  \includegraphics[width=.45\textwidth]{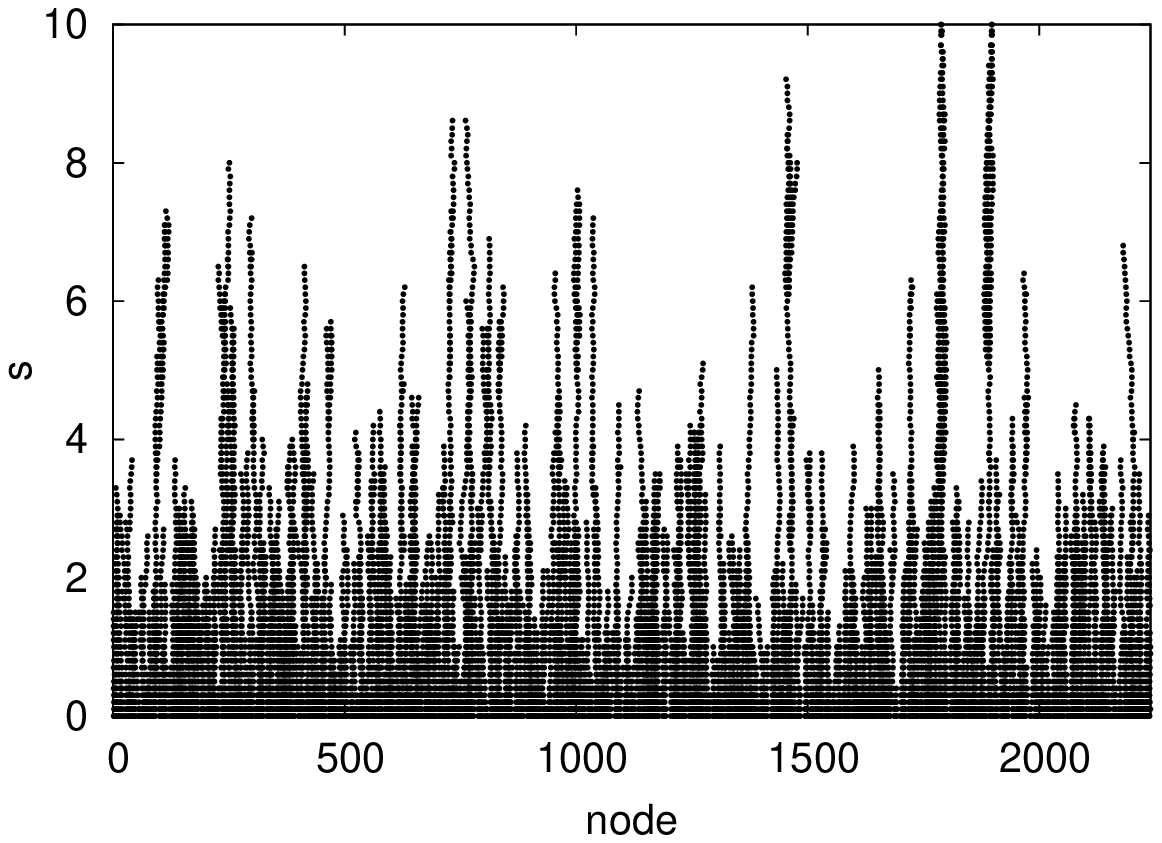}
  \includegraphics[width=.45\textwidth]{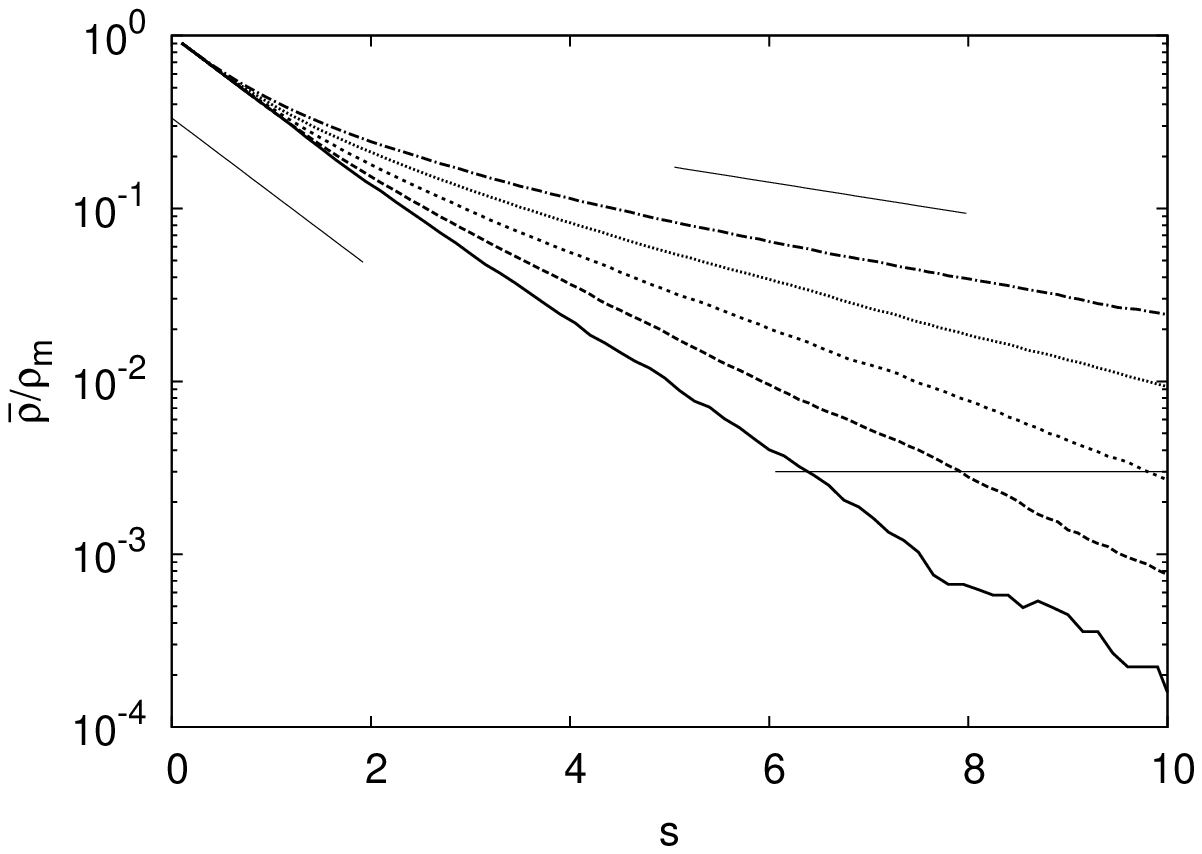}
  \caption{Left: spatiotemporal plot of a single realization of the particle dynamics (space in horizontal,
  time in vertical) for $c_1=c_2=1,\ c_4=5\cdot 10^{-4}$ (region $R_1$ of figure \ref{fig:phase}).
  \corr{The non--dimensional time is $s=r_d t$.} Right: time dependence of the
  density (averaged over $100$ realizations) for $c_2=1$, $c_4=5\cdot 10^{-4}$, and (from bottom to top)
  $c_1=\frac{2}{10},\frac{4}{10},\frac{6}{10},\frac{8}{10},\frac{10}{10}$ (\ehg{all in} region $R_1$ of figure
  \ref{fig:phase}). Lines are used to indicate different theoretical stages of the dynamics.%
  \label{fig:2}
  }
\end{figure}

For the region of parameters of the system given by
$c_1=2r_b/r_d\le 1$, the results in Sec. \ref{sec:4} identify
extinction as the only homogeneous stable solution of the
density equation. We expect then it to be the final outcome
also of the particle dynamics. This is confirmed by all
numerical simulations run for this region. The left plot of
figure \ref{fig:2} shows a typical evolution of the particle
system from an initial condition where all nodes are filled
(space is plotted in the horizontal axis, and time in the
vertical). Particle number decays and they become extinct at
long times. When increasing $c_1$ the decay process \corr{lasts} longer
and finally, at some critical $c_1^{**}$ an absorbing-type
phase transition occurs beyond which there is an active phase
with a non-vanishing number of particles. On general grounds,
and as occurring in the case without excluded volume
\cite{Lopez2007}, we expect this transition to be of the
directed percolation type. The continuous description leading
to the solution $\rho_1$ in \eqref{eq:36} predicts $c_1^{**}=1$,
but in the particle system $c_1^{**}$ is always larger.

A quantitative comparison with the continuum description
requires averaging over multiple realizations to obtain the
expected density via $\rho(x,s)=\rho_m\medio{s_i}$. Additional
statistics is gained by further averaging in space to obtain
$\bar\rho(s)\equiv\int dx \rho(x,s)/L$. Indeed, because of the
translational invariance of the system, we expect the average
density to be already nearly homogeneous $\rho(x,s)\approx
\bar\rho(s)$ and then described by the homogeneous dynamics given by Eq.
\eqref{eq:34}. The right panel of Fig. \ref{fig:2} shows the
time dependence of $\bar\rho$, obtained from the particle
simulations. Initially, for high densities such that
$\rho/\rho_m \ge \frac{c_1}{c_2}$, dynamics is dominated by
deaths and should agree with stage (i) in Sec. \ref{sec:4}, so
that $\bar\rho$ would have an exponential decay with rate $r_d$
(or $1$ in units of $r_d$, Eq. \eqref{eq:35}). Moreover, the
larger the birth rate $r_b$, keeping the rest of the parameters
constant, the smaller the time the system spends in this first
stage, in accordance with the estimation of $s_0$ just before
Eq. \eqref{eq:35}. For smaller densities a transient would
occur within stage (ii), but when $\bar\rho$ becomes
sufficiently small, the linear dynamics of Eq. \eqref{eq:34} or
equivalently, Eqs. \eqref{eq:37}-\eqref{eq:38} with $\vec
n=\vec 0$, should take over. Then we expect at long times an
exponential decay $\bar\rho\sim\exp(-|\lambda| s)$ with
exponent $|\lambda|=1-c_1$ in units of $r_d$ (Eq. \eqref{eq:38}
for $\vec n=\vec 0$). A third regime, corresponding to the
breakdown of the continuum limit (when $\rho/\rho_m \sim 1/N$),
also appears. It is characterized by the presence of large
fluctuations.

\begin{figure}[!h]
  \centering
  \includegraphics[width=.45\textwidth]{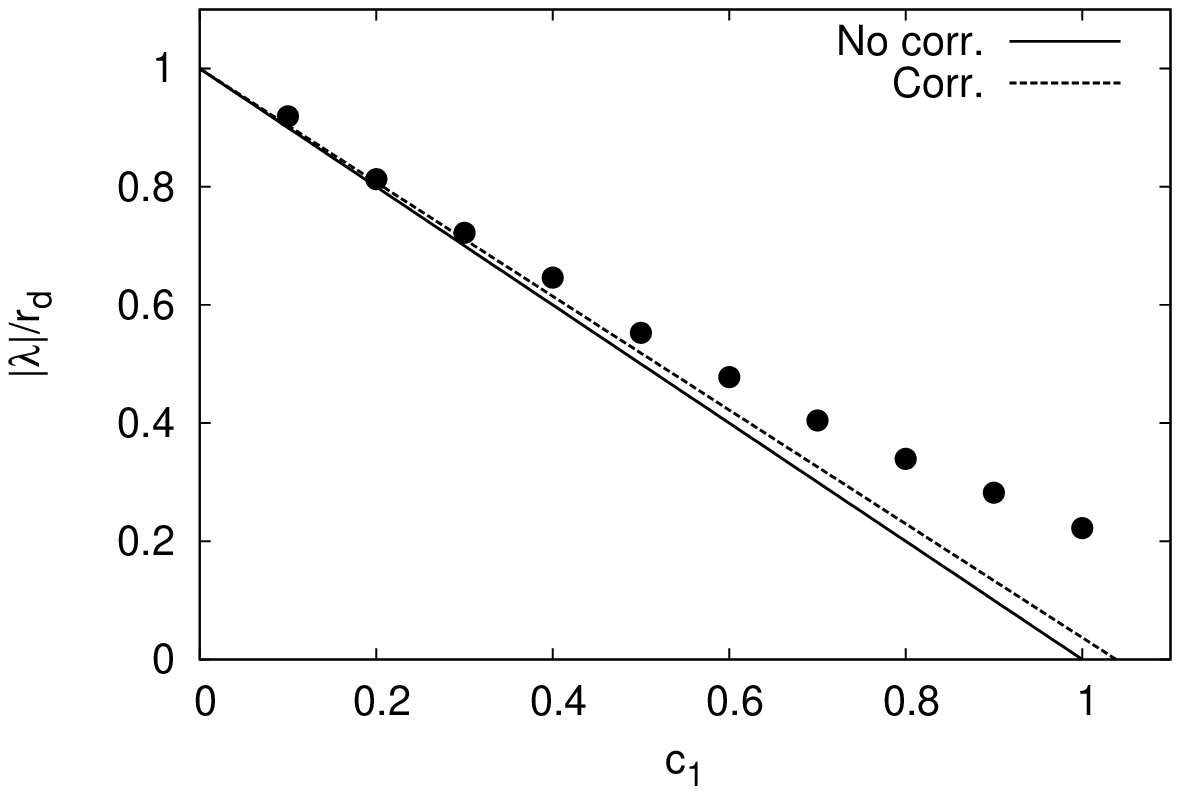}
  \includegraphics[width=.45\textwidth]{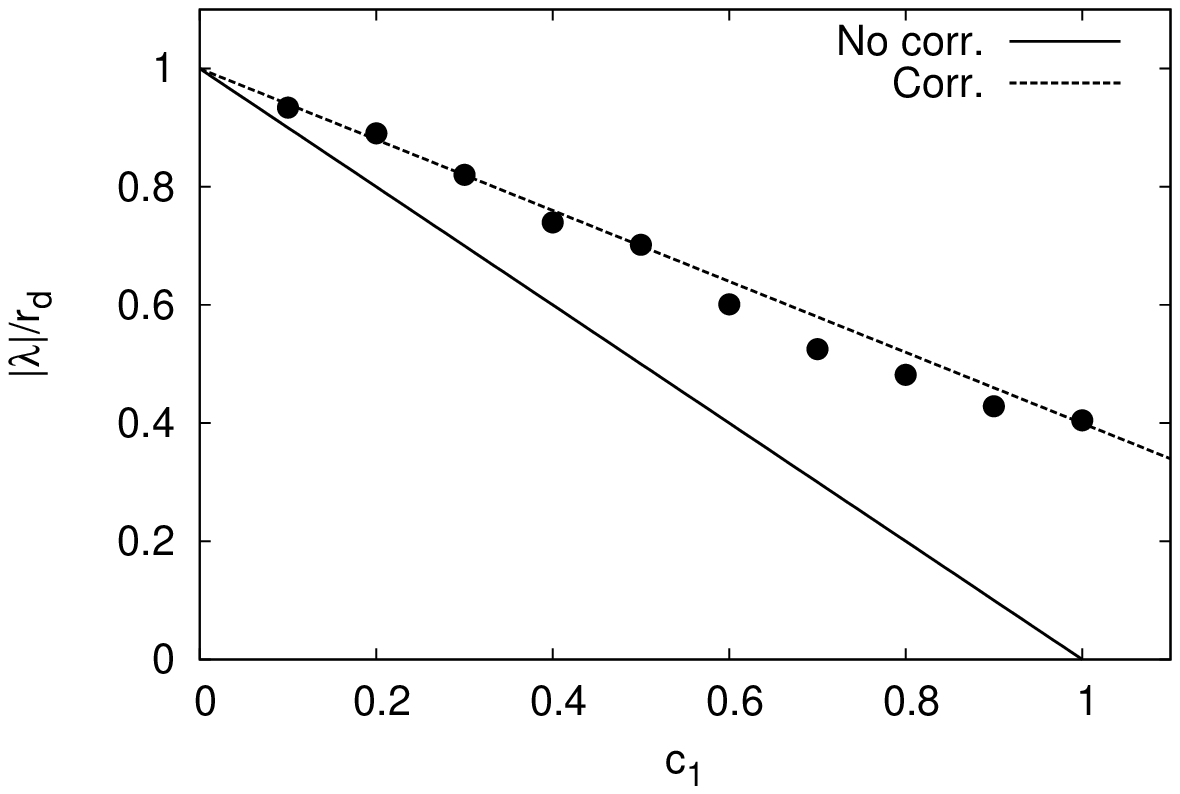}
  \caption{Decay rate $|\lambda|$ fitted to the exponential
decay towards zero of the mean density at long times, $\bar\rho \sim
\exp(-|\lambda| s)$, obtained as a function of $c_1$ from particle simulations
  (symbols). Left: $c_2=1$, $c_4=5\cdot 10^{-4}$.  Right: $c_2=1$, and $c_4=10^{-5}$.
  Solid lines give the linear prediction from Eq. \eqref{eq:28} (or from \eqref{eq:28} or \eqref{eq:34}),
  namely $|\lambda|/r_d=1-c_1$. The dashed lines give the corresponding decay rate from linearization of
  Eq. \eqref{eq:31}, which improves the approximation of the correlations in the dilute limit. 
  }
  \label{fig:time}
\end{figure}

Fig. \ref{fig:time} shows (symbols) the decay rates $|\lambda|$
fitted to the exponential decay towards zero of the mean
density at long times (but before the large-fluctuation final
regime). The two panels represent two different situations. In
the left one the mobility $r_m$ is larger so that factorization
of correlations is justified and the linearization of Eq.
\eqref{eq:28} (or from \eqref{eq:28} or \eqref{eq:34}), works
well for small $c_1$. When increasing $c_1$, however, the
agreement worsens. We attribute this to the increasing
fluctuations occurring when approaching the critical region of
the phase transition to the active phase, which occurs at
$c_1=1$ according to the continuum theory and at a larger value
in the particle system. Large critical fluctuations invalidate
the assumptions leading to our Eq. \eqref{eq:28}. For the right
panel in Fig. \ref{fig:time}, mobility is smaller and
nontrivial correlations are apparent at all values of $c_1$.

Since in the late stages of density decay $\rho$ is small, it
is likely that the approach in Sect. \ref{subsubsec:dilute}, in
which correlations are taken into account approximately in the
dilute situation, would work. We linearize Eq. \eqref{eq:31}
close to $\rho\approx 0$ and plot the resulting decay rate of
homogeneous perturbations as a dashed line in the two panels of
Fig. \ref{fig:time}. We observe some improved agreement,
specially in the small mobility (right) case. But still it
fails to describe completely the critical region.

\subsection{Large birth rates. Homogeneous and non-homogeneous states}

For $c_1\ge 1$, the theory of section \ref{subsubsec:factor}
predicts the existence of non-vanishing homogeneous density
solutions, given by Eq. \eqref{eq:37}. The new solutions have
nonzero steady-states values which are stable for $1\le c_1\le
c_1^*(c_2,c_4)$ and unstable for $c_1\ge c_1^*$. In region
$c_1\ge c_1^*$ the system would exhibit spatial patterns for
long times. For $d=1$, the critical surface $c_1^*(c_2,c_4)$ is
given by Eqs. \eqref{eq:42}-\eqref{eq:43}. We see (Fig.
\ref{fig:phase}) that for small $c_2$ the system should be
homogeneous, and for large $c_1$ and $c_2$ it should be in a
spatially periodic state.

Simulations show partial agreement with these predictions, but
also some discrepancies. First, as stated above, the transition
from the extinct to the active phase does not occur at $c_1=1$,
but at $c_1=c_1^{**}(c_2,c_4)> 1$ (see asterisks in Fig.
\ref{fig:phase}). For example, for $c_4=5\cdot 10^{-4}$, and
for both $c_2=1$ and $c_2=10$, $c_1^{**}\simeq 1.5$. \corr{This
can be understood \ehg{from a combination of} two factors: the
stabilization effect of the extinction solution induced by the
volume exclusion \ehg{(via the reduction of the effective birth
rate)}, and the neglected fluctuations and correlations
\ehg{(which are known to stabilize the extinct phase in this
type of absorbing transition \cite{lohe04,Lopez2007}).}  }

\begin{figure}[!h]
  \centering
  \includegraphics[width=.45\textwidth]{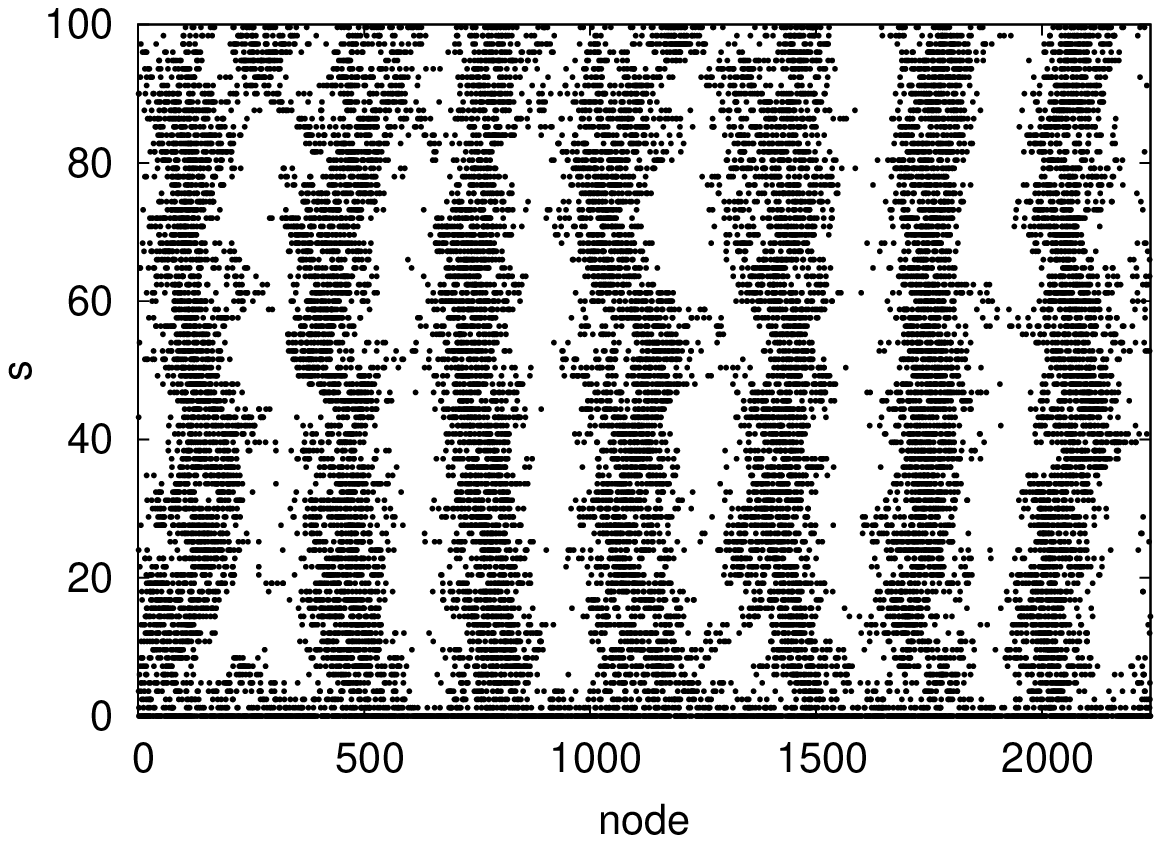}
  \includegraphics[width=.45\textwidth]{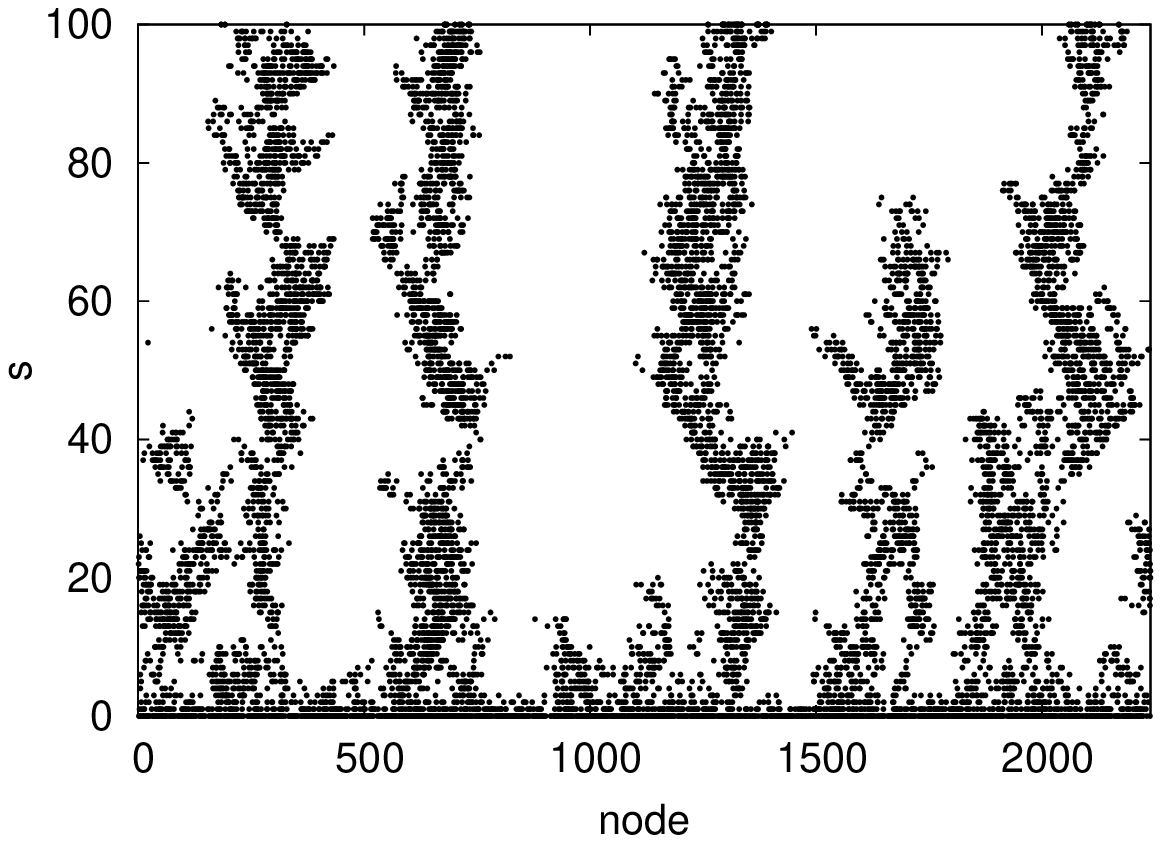}
  \caption{Left: Time evolution of a system with $c_1=3$, $c_2=20$, and $c_4=5\cdot 10^{-4}$. Right:
  evolution of a system with $c_1=1.7$, $c_2=10$, and $c_4=5\cdot 10^{-4}$.
  \corr{The non--dimensional time is $s=r_d t$.}  }
  \label{fig:4}
\end{figure}

Second, in agreement with the theory, there are periodic (see
Fig. \ref{fig:4}, left) and homogeneous (not shown) active
phases, occurring roughly in the regions predicted, except for
the absence of any active phase in the region
$1<c_1<c_1^*(c_2,c_4)$ (see Fig. \ref{fig:phase}). The spatial
periodicity of the state in the left panel of Fig. \ref{fig:4},
is (in units of $L$) approximately $l\approx 1.43 R \approx
0.143$, in agreement with the rule $l\in [R,2R]$ which arises
from general competitive \ehg{interaction} considerations
\cite{hehelo15} and from identification of the fastest growing
Fourier mode in the linear stability analysis of Sect.
\ref{sec:4}.  \ehg{This theoretical Fourier mode corresponding
to the nearest critical point located at $c_1=10$ and
$c_2\simeq 1.5$ gives $\gamma_n^c\approx 4.3$ from which the
predicted periodicity is $l^c\simeq 1.46 R$, very close to the
observed one. In fact $l\approx 1.43 R$ is the periodicity
allowed by the periodic boundary conditions closest to $l^c$}.

Third, the continuum theory predicts that the sequence of
steady states encountered when increasing $c_1$ at a
large-enough value of $c_2$, from the zero solution to periodic
patterns, includes homogeneous states. The particle simulations
do not show this feature, but a direct transition from
extinction to inhomogeneous states occurs instead. Whereas the
inhomogeneous states develop a clear spatial periodicity when
sufficiently far from the transition region (Fig. \ref{fig:4},
left), close to the transition line they are more irregular and
strongly fluctuating (Fig. \ref{fig:4}, right).

\begin{figure}[!h]
  \centering
  \includegraphics[width=.7\textwidth]{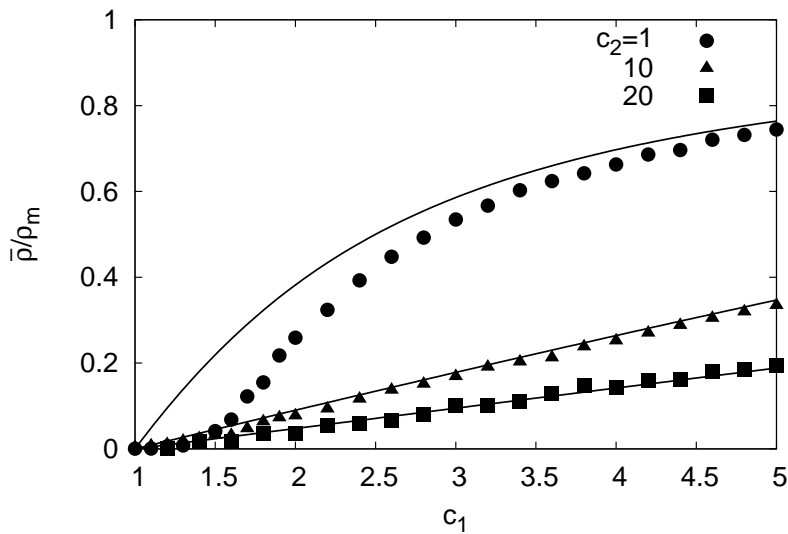}
  \caption{Steady values of $\bar\rho/\rho_m$ as a function of $c_1$ for $c_4=5\cdot 10^{-4}$
  and $c_2=1$ (circles) and $c_2=10$ (triangles). Points correspond to simulations (averaged over
  $10$ realizations) and lines to the theory. 
  }
  \label{fig:5}
\end{figure}

We compare in Fig. \ref{fig:5} the mean density $\bar\rho$ in
the steady state of the particle system with the homogeneous
solution $\rho_1$ (Eq. \eqref{eq:36}) of Eq. \eqref{eq:34} for
several values of $c_2$. For $c_2=1$ the particle
configurations are statistically homogeneous and then we
expect, and observe, reasonable agreement sufficiently far
apart from the critical transition region. Fluctuations shift
the appearance of the active phase from the deterministic
$c_1=1$ value to the observed $c_1\approx 1.5$. Close to this
transition the particle density is smaller than predicted, as
usual in similar absorbing-transition situations
\cite{lohe04,Lopez2007}.  For $c_2=10$ and $20$ both the particle system
and the theoretical description agree in which translational
symmetry is broken and inhomogeneous configurations appear.
Nevertheless, mean density is quite small, so that coupling of
the different modes with the homogeneous one is not strong and
still we have that the mean density is well described by the
homogeneous prediction: $\bar\rho\approx \rho_1$. 

\section{Discussion \label{sec:6}}

We have studied the effects of volume exclusion on a
birth-death model with spatially non-local rates. The system
models competition since the reproduction rate of any particle
decreases if the number of other particles increases in the
spatial surroundings.
The stochastic model is the basis for macroscopic descriptions
and of the simulation results. By taking moments of the master
equation of the stochastic model, we encountered an infinite
hierarchy of equations. To make the latter description useful,
we closed this hierarchy by proposing approximations, some of
them beyond the usual correlation factorization approximation
and passed to the continuum limit.

Disregarding the volume of the particles we recover previous
continuum equations for the density of the system. If the
volume of particles is considered, the description changes by
the appearance of a new factor in the birth term,
$1-\rho/\rho_m$. This is given in equation \eqref{eq:28} which
is a central result of our work. The effect of the new factor
is twofold. First, birth rate is always reduced, so that
generally the system with volume exclusion will exhibit smaller
densities than without it. Second, because of the appearance of
two qualitatively new terms in the density equation
\eqref{eq:28} compared to the case without volume exclusion,
Eq. \eqref{eq:23}, namely the ones proportional to $\rho^2$ and
to the cubic $\rho^2\int\rho~dr$, there are two additional
dimensionless parameters governing the system behavior, which
makes the phase diagram (Fig. \ref{fig:phase}) richer. In
particular regions where the continuum theory predicts
homogeneous non-vanishing density is extended towards large
$c_1$ when $c_2$ is small, whereas it only appears for small
$c_1$ in the absence of volume exclusion \cite{lohe04}. The
steady homogeneous solutions of \eqref{eq:28} have been fully
characterized and the linear and homogeneous nonlinear dynamics
towards them discussed. Comparison with the numerical
simulations reveals the \corr{good agreement} of the
description for some situations, but at the same time the
importance of the correlations \corr{when the mobility is small
enough (e.g. Fig. \ref{fig:time}) and/or near the critical
points separating regions $R_1$ and $R_2$ of figure
\ref{fig:phase}. For bigger values of $r_m/r_d$, keeping the
rest of the parameter constant, the factorization of the
correlation as the square of the density becomes a better
approximation, and better agreement is obtained. The chosen
values of the parameters allowed \ehg{us} to determine the
\ehg{limitations} of equation \eqref{eq:28}}.

Linear stability analysis reveals the presence of periodic
solutions for \corr{large} birth and competition, and small mobility.
The particle simulations also display these states, although
they become rather fluctuating and irregular close to
transition points.

\corr{The model proposed in this work can be naturally extended
along different directions. Regarding the geometry of the
underlying lattice, it can be chosen to be a complex network,
in general. This would imply direct changes of the different
terms appearing on the macroscopic equations. Since the death
rate depends on the number of particles and not on the form
particles are arranged, its associated terms would not change
upon modifying the lattice. Furthermore, at least at the
leading order considered in this work where \ehg{immediate}
diffusion of particles coming from the birth term is neglected,
all geometry dependence of births is encoded in the correlation
and the density functions, and hence the influence of the
lattice is minimal. Finally, the geometry of the underlining
network \ehg{would be important upon obtaining the result of a
diffusion term with position--independent diffusion
coefficient, which in the case discussed here of an hypercubic
lattice in arbitrary dimension requires only of the restriction
on the number of allowed particles per node to be $\sigma=1$ or
$\sigma=\infty$}. In general, terms that account for movement
of particles are different for different lattices. We can go
further by allowing the underlying lattice to deform or move in
a continuum space, in such a way that it describes now the
collective motion of particles like in a crowded environment,
extending previous works \cite{biyamama16}.}

We close this discussion by commenting that consideration of
volume exclusion opens the door to the study of a large, rather
unexplored, field in spatial population dynamics: mutualistic
systems. They are characterized by the increase (\corr{respectively} decrease) of
reproduction (death) rates due to the presence of other
individuals. Without spatial-dependence most of the models are
ill-posed since density grows without limit (see however
\cite{Holland2002}). In our spatially-dependent case, mutualism
would correspond to considering $\alpha$ negative. Due to the
excluded volume effect, particle density cannot grow
indefinitely. Thus, our model and equations, taken for
$\alpha<0$ can be considered to be a well-posed framework to
study mutualistic interactions. A detailed study of this case
will be presented in future work.


\ack We acknowledge financial support \ehg{from grants LAOP,
CTM2015-66407-P (AEI/FEDER, EU) and ESOTECOS,
FIS2015-63628-C2-1-R and FIS2015-63628-C2-2-R (AEI/FEDER, EU).}

\end{document}